# ISOTOPIC TABLES OF THE BARYONS AND MESONS (2003)

David Akers*
*Lockheed Martin Corporation, Dept. 6F2P, Bldg. 660, Mail Zone 6620,*
1011 Lockheed Way, Palmdale, CA 93599
*Email address: David.Akers@lmco.com

**Abstract**

A constituent-quark (CQ) mapping of all the baryon and meson resonances is listed for < 3000 MeV. The *Isotopic Tables of the Baryons and Mesons (2003)* are based upon the CQ Model of M. H. Mac Gregor. Hadrons in boldface are predictions of Mac Gregor. Hadrons in blue indicate rotational states. Predicted hadrons and their spins are indicated in red color. Underlined baryons indicate SU(3) decuplet for J = 3/2. The tables may be extended to charmed and bottom baryons or mesons at a later date. The table for the baryons is expanded to incorporate the recently discovered S = + 1 baryon. Additional rotational states of baryons are predicted.

## INTRODUCTION

In the *Isotopic Table of the Baryons (2003)*, we list all the well-established baryons for < 3000 MeV from the Particle Data Group [1]. The not-so-well-established baryon resonances are also included in the table. From left to right in the table, the vertical columns represent increasing energy in quantized units of mass m = 70 MeV. From top to bottom, the horizontal rows also represent increasing energy in quantized units of m = 70 MeV. Each baryon is grouped according to its mass and angular momentum.

For the Δ baryon resonances, Mac Gregor utilized the nucleon N (939) as the ground state [2]. By the selection of the N (939) core for the Δ baryons, there is an inconsistency with the Gell-Mann quark model in which the SU(3) decuplet for J = 3/2 has equally spaced mass separations of approximately 140 MeV. Therefore, we make core corrections for the Δ baryons [3]. Likewise, we note that the core selection by Mac Gregor for the Ω baryons was incorrect for the same reasons as stated in regards to satisfactorily explaining the masses of the SU(3) decuplet. Mac Gregor chose to list the Δ baryons after the (Δ - N) core correction and chose to place the Ω(1672) in the ground state with J = ½. However, Ω(1672) has a J = 3/2 and should be located in the *same* CQ excitation band as the other members of the SU(3) decuplet. The SU(3) decuplet is noted as the underlined baryons in row F(210). Thus, there must exist a Ω(1499) core with spin J = ½ which is not previously known and which is shown in the table.

The Δ(1232)P is now located in row F(210) for the reasons previously explained. The baryons in row F(210) all have spin J = 3/2 in the P-state. This establishes a consistency with the Gell-Mann quark model. However, there remains what to select for the Δ baryon core. It is obvious from the ground states across all isospins that the Δ core must have J = ½ in the P-state and that it must have a mass of about 1079 MeV. With the Δ(1079)P core correction, we introduce a new column of Δ baryons in the table, where all baryons have the *same* J-value in each row. Moreover, we have a new column of Ω baryons as well. In the table, the baryons in boldface are predictions of Mac Gregor [4-5]. Those baryons in blue color indicate rotational states [6]. Newly predicted baryons and their spins are indicated in red color [3].

## S = + 1 BARYONS

In 2003, several collaborations have discovered an exotic baryon with five quarks [7-10]. The recently discovered exotic baryon has been named θ$^{+}$(1540). This baryon was a seminal prediction by Diakonov, Petrov, and Polyakov [11]. These authors utilized a Skyrme-inspired quark soliton model. The idea that many baryons are rotational excitations originated with Mac Gregor [4]. In studying the rotational spectra of hadrons, Mac Gregor noted several regularities in the mass systematics [6]. The *Isotopic Table of the Baryons* is based upon these regularities of particle masses [2, 3]. The table is expanded to incorporate the recently discover $\theta^+$ (1540).

In the paper by Diakonov, Petrov and Polyakov [11], there are predictions of particle masses for the anti-decuplet. These authors associate $\theta^+$ (1530) with N(1710), $\Sigma$(1890), and $\Xi$(2070) as shown in Fig. 1. Jaffe and Wilczek predict exotic $\Xi$ baryons to be 300 MeV lighter and suggest that $\theta^+$ (1530) may be associated with N(1440) [12]. On the other hand, Glozman believes that it is incompatible to associate $\theta^+$ (1530) with the Roper resonance N(1440), because of the very narrow width of $\theta^+$ (1530) compared to the very broad widths of N(1440), $\Lambda$(1660), and $\Sigma$(1660) [13]. From the *Isotopic Table of the Baryons*, we calculate particle masses for the anti-decuplet. Our masses are slightly *lower* than the original predictions of these authors and are shown in Fig. 2. It remains to be seen which nucleon is associated with the new S = + 1 baryon $\theta^+$ (1540), and therefore its position in the table is tentative.

## MESONS

In the *Isotopic Table of the Mesons (2003)*, we list all the well-established mesons for < 2700 MeV from the Particle Data Group [1]. The not-so-well-established meson resonances are also included in the table. From left to right in the table, the vertical columns represent increasing energy in quantized units of mass m = 70 MeV. From top to bottom, the horizontal rows also represent increasing energy in quantized units of m =

70 MeV. Each meson is grouped according to its mass *only* for the current configuration. A sort on angular momentum may be considered at a later date.

In the table, the mesons are grouped according to the scheme and notation of the constituent-quark (CQ) Model from M. H. Mac Gregor [2]. The mesons in boldface are unobserved levels from Mac Gregor [4]. Newly predicted mesons and their spins are indicated in red color [14]. These predictions are based upon a magnetic monopole model of hadrons. Thus, their positions are tentative in the table.

In contrast to the prediction of η(1820) in Ref. 14, the other scalar mesons are tentatively accepted as σ(560) and κ(900). From a study of Regge trajectories [15], we predict another scalar $f_0(850)$. The CQ Model is a purely electric coupling model of quarks without magnetic charges. In the CQ Model, many of the meson and baryon states are rotational levels [6]. These rotational levels may be indicated in the table at a later date. Particle widths are indicated in boldface in the lower left for each box (in MeV).

As a final note about the *Isotopic Table of the Mesons*, we have calculated core corrections for the mesons. For each isospin, core masses from the ground states are subtracted from each meson resonance in the vertical column of the table. The differences are then plotted in Fig. 3. In Fig. 3, the green colored levels are I = 1 meson differences with respect to π (135). The magenta colored levels represent I = ½ meson differences with respect to K (494), and the blue colored levels indicate I = 0 levels with respect to η (547).

In Fig. 3, it can be easily noted that there are patterns of energy separations: F = 210 MeV, BF = 350 MeV, and X = 420 MeV. This notation comes from the CQ Model of Mac Gregor [2]. Similar patterns of energy separations, m = 70 MeV and F = 210 MeV, are found for the baryons [3]. Since the energy differences are *experimental* values, the patterns of energy separations will never go away.

## ACKNOWLEDGEMENT

The author encourages comments, suggestions and any corrections from readers of the *Isotopic Tables of the Baryons and Mesons*. The author wishes to thank Dr. Malcolm Mac Gregor, retired from the University of California's Lawrence Livermore National Laboratory, for his encouragement to pursue the CQ Model, and he wishes to thank Dr. Paolo Palazzi of CERN for his interest in the work and for e-mail correspondence.

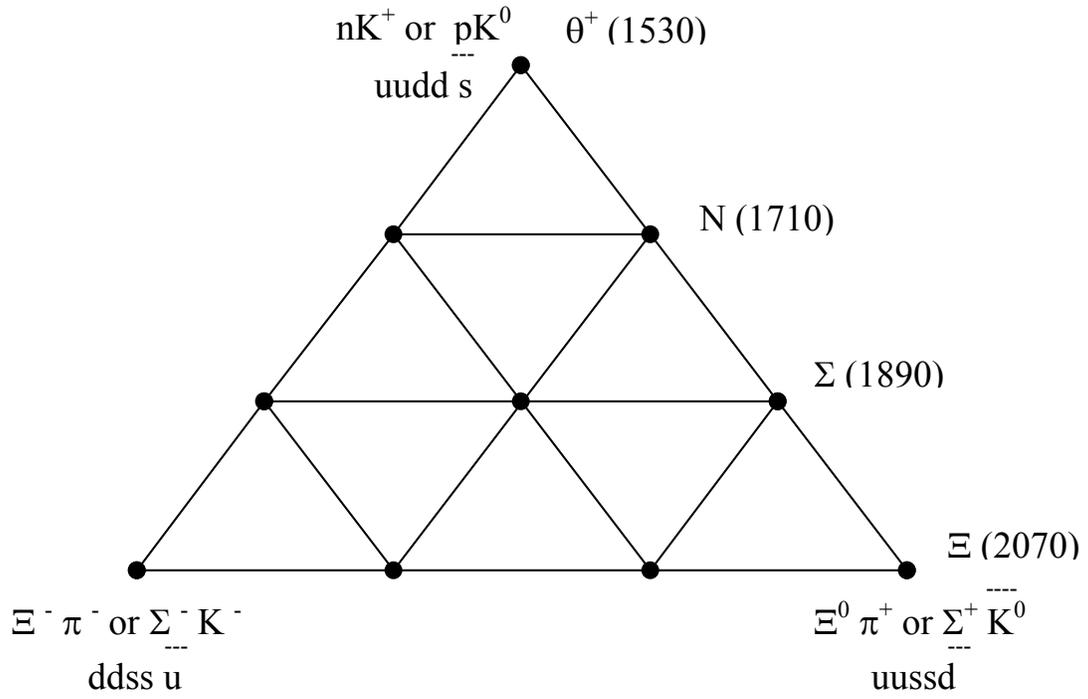

Fig. 1. The suggested anti-decuplet of baryons [11].

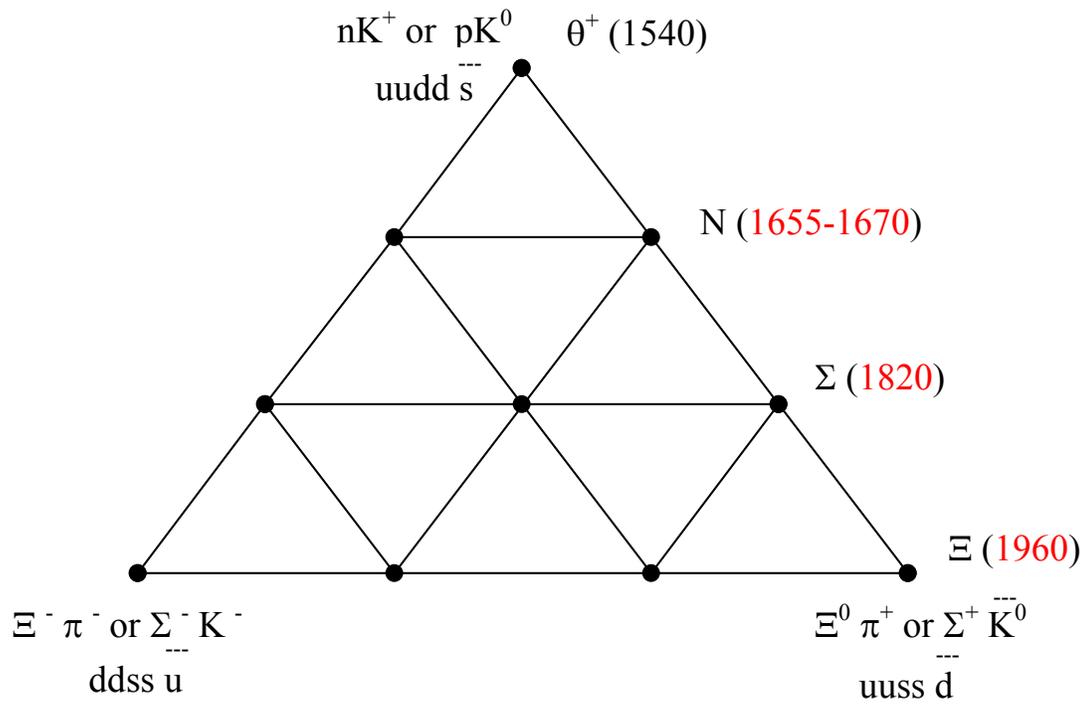

Fig. 2. The predicted masses from the CQ Model.

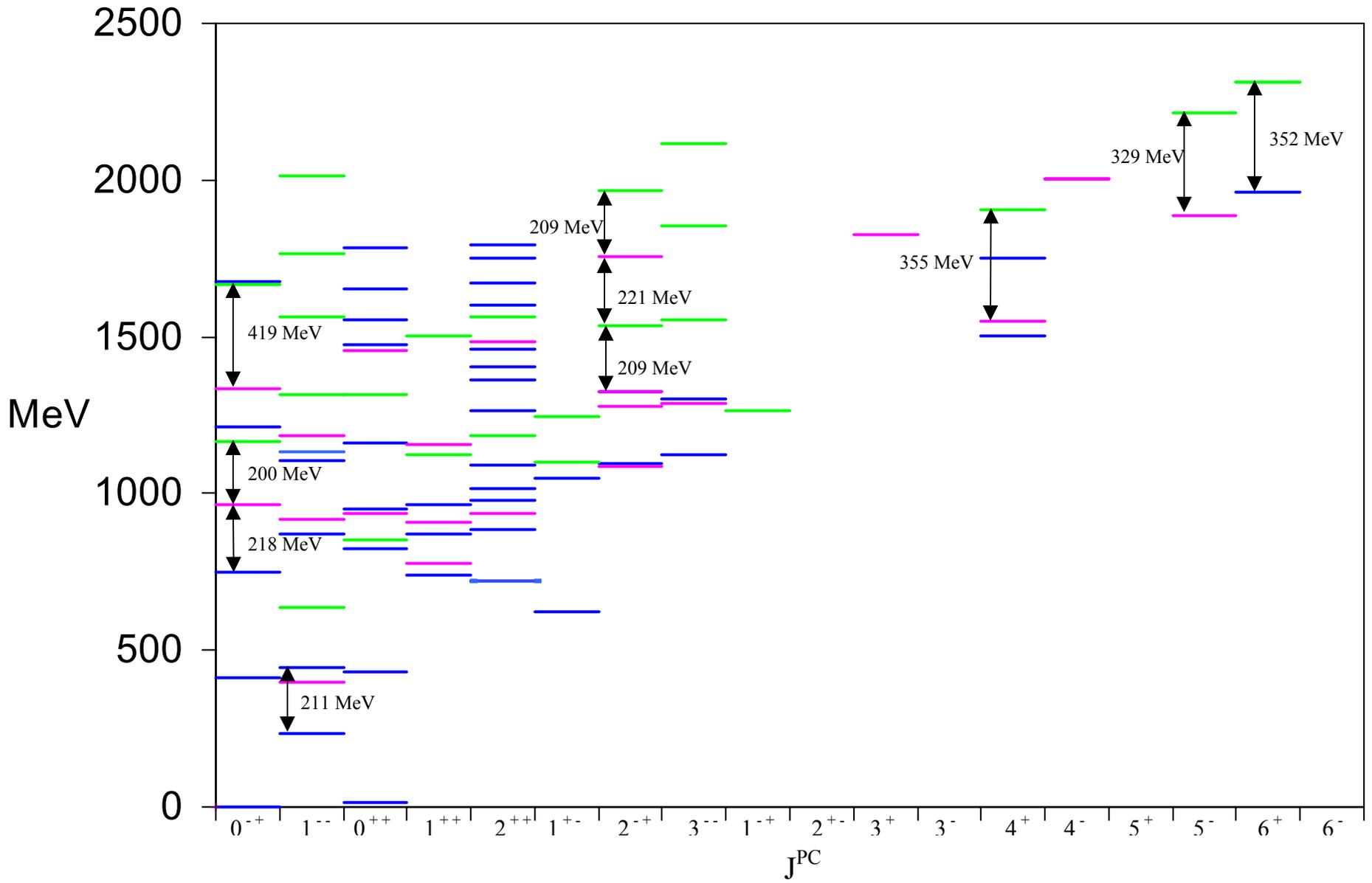

Fig. 3

# ISOTOPIC TABLE OF THE BARYONS (2003)

A constituent-quark (CQ) mapping of all the baryon resonances is listed for < 3000 MeV. Table is based upon the CQ Model of Malcolm H. Mac Gregor (1990). Baryons in boldface are predictions of Mac Gregor (1974). Baryons in blue indicate rotational states. Predicted baryons and their spins are indicated in red color. Underlined baryons indicate SU(3) decuplet for J = 3/2. Table may be extended to charmed and bottom baryons at a later date. Table expanded to incorporate recently discovered S = + 1 baryon. Particle width is indicated in lower left for each box (in MeV).

| CQ Shell Numbers | N – 2m | N | N + 2m | N + 3m | N + 4m | N + 6m | N + 8m |
|---|---|---|---|---|---|---|---|
| Strangeness | +1 | 0 | 0 | -1 | -1 | -2 | -3 |
| Isospin |  | 1/2 | 3/2 | 0 | 1 | 1/2 | 0 |
| Ground State (in MeV) |  | N<br>N(939)$P_{11}$ | Δ = NB<br>Δ(1079)$P_{31}$ | Λ = NF<br>Λ(1116)$P_{01}$ | Σ = NBB<br>Σ(1192)$P_{11}$ | Ξ = NBBB<br>Ξ(1321)$P_{11}$ | Ω = NBBBB<br>Ω(1499)$P_{01}$ |
| CQ excitation (in MeV) |  |  |  |  |  |  |  |
| m(70) |  |  | **Δ(1149)$S_{31}$** |  | Σ(1262)$S_{11}$ | Ξ(1391)$S_{11}$ | Ω(1570)$S_{01}$ |
| B(140) |  |  |  |  |  |  |  |
| F(210) |  |  | <u>Δ(1232)$P_{33}$</u><br>120 |  | <u>Σ(1385)$P_{13}$</u><br>35.8 | <u>Ξ(1530)$P_{13}$</u><br>9.9 | <u>Ω(1672)$P_{03}$</u> |

| | | | | | | |
|---|---|---|---|---|---|---|
| | | **N(1149)S$_{11}$** | | Λ(1326)S$_{01}$ | | |
| BB(280) | | **N(1219)S$_{11}$** | | Λ(1405)S$_{01}$<br>**50±2** | Σ(1480)S$_{11}$<br>**80±20** | Ξ(1620)S$_{11}$<br>**40±15** | |
| FB(350) | | **N(1300)P$_{11}$** | | **Λ(1460)P$_{01}$** | **Σ(1560)P$_{11}$**<br>**79±30** | | |
| | | **N(1343)D$_{13}$** | | **Λ(1520)D$_{03}$**<br>**15.6** | **Σ(1580)D$_{13}$**<br>**15** | **Ξ(1690)D$_{13}$**<br>**44±23** | **Ω(1868)D$_{03}$** |
| X(420) | | **N(1359)S$_{11}$** | **Δ(1540)S$_{31}$** | **Λ(1536)S$_{01}$** | Σ(1620)S$_{11}$<br>**87±19** | **Ξ(1741)S$_{11}$** | |
| FBB(490) | | **N(1440)P$_{11}$**<br>**350** | | **Λ(1600)P$_{01}$**<br>**250** | **Σ(1660)P$_{11}$**<br>**200** | **Ξ(1800)P$_{11}$** | |
| | | **N(1462)D$_{13}$** | | **Λ(1600)D$_{03}$** | **Σ(1670)D$_{13}$**<br>**60** | **Ξ(1820)D$_{13}$**<br>**24±15** | **Ω(1962)D$_{03}$** |
| | | | | | **Σ(1690)P$_{13}$**<br>**240±60** | | |
| | | | **Δ(1600)P$_{33}$**<br>**350** | | | | |
| BX(560) | | | Δ(1620)S$_{31}$<br>**150** | Λ(1670)S$_{01}$<br>**35** | Σ(1750)S$_{11}$<br>**90** | | **Ω(2070)S$_{01}$** |

| | | | | | Σ(1770)P$_{11}$<br>80±30 | | |
|---|---|---|---|---|---|---|---|
| | | | Δ(1700)D$_{33}$<br>600±250 | | | | |
| | | | | | Σ(1775)D$_{15}$<br>120 | | |
| | | N(1520)D$_{13}$<br>120 | | Λ(1690)D$_{03}$<br>60 | Σ(1780)D$_{13}$ | Ξ(1950)D$_{13}$<br>60±20 | |
| | | N(1535)S$_{11}$<br>150 | | | | | |
| FX(630) | | N(1580)P$_{13}$ | | | Σ(1840)P$_{13}$<br>120±30 | | |
| | | | Δ(1750)P$_{31}$<br>300±120 | | | | |
| BXB(700) | | N(1650)S$_{11}$<br>150 | | Λ(1800)S$_{01}$<br>300 | | | |
| | | N(1670)P$_{11}$ | | Λ(1810)P$_{01}$<br>150 | Σ(1880)P$_{11}$<br>80 to 260 | Ξ(2025)P$_{11}$ | |
| | | N(1680)F$_{15}$<br>130 | | Λ(1820)F$_{05}$<br>80 | Σ(1915)F$_{15}$<br>120 | Ξ(2030)F$_{15}$<br>20±15 | |

| | | N | Δ | Λ | Σ | Ξ | Ω |
|---|---|---|---|---|---|---|---|
| | | N(1675)D$_{15}$ 150 | **Δ(1800)D$_{35}$** | **Λ(1830)D$_{05}$** 95 | | | |
| FBX(770) | | **N(1700)D$_{13}$** 100 | | | **Σ(1940)D$_{13}$** 150 to 300 | | |
| | θ$^+$(1540) | **N(1710)P$_{11}$** 150 | | **Λ(1890)P$_{01}$** | **Σ(1960)P$_{11}$** | **Ξ(2090)P$_{11}$** | |
| | | | **Δ(1870)D$_{33}$** | | | **Ξ(2120)D$_{13}$** 25±12 | **Ω(2250)D$_{03}$** 55±18 |
| | | | | | Σ(2000)S$_{11}$ 116 to 450 | | |
| | | N(1720)P$_{13}$ 150 | | **Λ(1890)P$_{03}$** 100 | | | |
| XX(840) | | | Δ(1900)S$_{31}$ 200 | | | | |
| | | | | | **Σ(2030)F$_{17}$** 150 to 200 | | |
| | | | **Δ(1905)F$_{35}$** 400±100 | | **Σ(2070)F$_{15}$** 300 | | |
| | | | **Δ(1910)P$_{31}$** 250 | | | | |
| | | | Δ(1920)P$_{33}$<br>**200** | | **Σ(2080)P$_{13}$**<br>**240** | | |
| | | | **Δ(1930)D$_{35}$**<br>**530±140** | | | | |
| | | **N(1830)D$_{13}$** | Δ(1940)D$_{33}$<br>**460±320** | | | | |
| | | | **Δ(1950)F$_{37}$**<br>**300** | | **Σ(2100)G$_{17}$**<br>**135±30** | **Ξ(2250)G$_{17}$**<br>**46±27** | **Ω(2384)G$_{07}$**<br>**26±23** |
| | | | | **Λ(2000)G$_{07}$**<br>**180 to 240** | | | |
| mXX(910) | | | **Δ(2000)F$_{35}$**<br>**250** | | | | |
| | | **N(1880)F$_{15}$** | | | | | |
| | | | | **Λ(2020)F$_{07}$**<br>**160** | | | |
| | | **N(1900)P$_{13}$**<br>**498±78** | | | | | |
| BXX(980) | | **N(1910)G$_{17}$** | | **Λ(2100)G$_{07}$**<br>**200** | | | **Ω(2470)G$_{07}$**<br>**72±33** |
| | | | | $\Lambda(2110)F_{05}$ **200** | | $\Xi(2370)F_{15}$ **80±25** | |
| FXX(1050) | | N(1990)F$_{17}$ **535±120** | | | | | |
| | | | $\Delta(2140)G_{37}$ | | $\Sigma(2250)G_{17}$ **100** | | |
| | | **N(2000)F$_{15}$** **490±310** | | | | | |
| | | | $\Delta(2150)S_{31}$ **200±100** | | | | |
| B$_2$X$_2$(1120) | | **N(2080)D$_{13}$** **450±185** | | | | $\Xi(2500)D_{13}$ **150±60** | |
| | | N(2090)S$_{11}$ **414±157** | | | | | |
| | | **N(2100)P$_{11}$** **113±44** | | | | | |
| | | | **$\Delta(2200)G_{37}$** **450±100** | | | | |
| mB$_2$X$_2$(1190) | | | | **$\Lambda(2325)D_{03}$** **177±40** | | | |

| | | | Δ(2300)H$_{39}$<br>425±150 | Λ(2350)H$_{09}$<br>150 | | | |
|---|---|---|---|---|---|---|---|
| XXX(1260) | | N(2190)G$_{17}$<br>500±150 | | | | | |
| | | N(2200)D$_{15}$<br>400±100 | | | | | |
| | | N(2220)H$_{19}$<br>500±150 | | | Σ(2455)H$_{19}$<br>140 | | |
| | | | Δ(2350)D$_{35}$<br>400±150 | | | | |
| | | | Δ(2390)F$_{37}$<br>300±100 | | | | |
| mX$_3$(1330) | | N(2250)G$_{19}$<br>480±120 | Δ(2400)G$_{39}$<br>480±100 | | | | |
| | | | Δ(2420)H$_{3,11}$<br>450±150 | | | | |
| BX$_3$(1400) | | | Δ(2520)H$_{39}$ | | Σ(2620)H$_{19}$<br>221±81 | | |
| FX$_3$(1470) | | N(2420)H$_{19}$ | | Λ(2585)?$_{0?}$<br>150 | | | |

| FX$_3$B(1610) |  | **N(2550)H$_{19}$** |  |  |  |  |  |
|---|---|---|---|---|---|---|---|
| X$_4$(1680) |  | **N(2600)I$_{1,11}$** <br> 650 |  |  |  |  |  |
|  |  | **N(2700)K$_{1,13}$** <br> 900±150 |  |  |  |  |  |
|  |  |  | **Δ(2750)I$_{3,13}$** <br> 500±100 |  |  |  |  |
| BX$_4$(1820) |  |  | **Δ(2950)K$_{3,15}$** <br> 700±200 |  |  |  |  |
|  |  |  |  |  |  |  |  |

# ISOTOPIC TABLE OF THE MESONS (2003)

A constituent-quark (CQ) mapping of all the meson resonances is listed for < 2700 MeV. Table is based upon the CQ Model of Malcolm H. Mac Gregor (1990). Mesons in boldface are unobserved levels from Mac Gregor (1974). Predicted mesons and their spins are indicated in red color. Table may be extended to charmed and bottom mesons at a later date. Particle width is indicated in lower left for each box (in MeV).

| CQ Shell Numbers | | 2m | | 7m | | 8m |
|---|---|---|---|---|---|---|
| Isospin | 2 | 1 | | 1/2 | | 0 |
| Ground States (in MeV) | | $\pi^0(134.98)0^{-+}$ | | $K^{\pm}(494)0^-$ | | $\eta(547)0^{-+}$ <br> 0.00118 |
| | | $\pi^{\pm}(139.5)0^-$ | | $K^0(498)0^-$ | | **$f_0(560)0^{++}$** |
| CQ excitation (in MeV) | | | | | | |
| m(70) | | | | | | |
| B(140) | | | | | | |
| F(210) | | **M(350)** | | **K(704)** | | $\omega(782)1^{--}$ <br> 8.44 |
| BB(280) | | | | K(784) | | **f$_0$(850)** |
| FB(350) | | | | K*(892)1$^-$ **50.8±0.9** | | |
| X(420) | | | | | | η'(958)0$^{-+}$ **0.30** |
| | | | | | | f$_0$(980)0$^{++}$ **40 to 100** |
| FBB(490) | | | | | | φ(1020)1$^{--}$ **4.26** |
| BX(560) | | ππ(660) | | | | |
| FX(630) | | ρ(770)1$^-$ **149.2±0.7** | | | | h$_1$(1170)1$^{+-}$ **360±40** |
| BXB(700) | | | | | | f$_2$(1270) **184.3** |
| | | | | | | f$_1$(1285) **24.0±1.2** |
| | | | | | | η(1295) **55±5** |
| FBX(770) | | | | K$_1$(1270)1$^+$ **90±20** | | |
| XX(840) | | a$_0$(985)0$^{++}$ **50 to 100** | | | | f$_0$(1370) **255±60** |
| | | | | | | ω(1420) **174±59** |
| | | | | | | f$_1$(1420) **55.5±2.9** |
| | | | | | | f$_2$(1430) |
| | | | | | | η(1440) **56±6** |
| mXX(910) | | | | K$_1$(1400) **174±13** | | f$_0$(1500) **109±7** |
| | | | | K*(1410) **232±21** | | |
| | | | | K$_0$*(1430) **294±23** | | |
| | | | | K$_2$*(1430) **98.5±2.9** | | |
| BXX(980) | | | | K(1460) | | f$_1$(1510) **73±25** |

|  |  |  |  |  |  | f'$_2$(1525) 76±10 |
|---|---|---|---|---|---|---|
|  |  |  |  |  |  | f$_2$(1565) 126±12 |
| FXX(1050) |  |  |  | **K(1543)** |  |  |
|  | X(1600) 400±200 |  |  | K$_2$(1580) |  | h$_1$(1595) 384±100 |
|  |  |  |  |  |  | f$_2$(1640) 98±28 |
|  |  |  |  |  |  | η$_2$(1645) 181±11 |
|  |  |  |  |  |  | ω(1650) 220±35 |
| B$_2$X$_2$(1120) |  | b$_1$(1235) 142±9 |  | K(1630) 16±16 |  | ω$_3$(1670) 168±10 |
|  |  | a$_1$(1260) 250 to 600 |  | K$_1$(1650) 150±50 |  | φ(1680) 150±50 |
|  |  |  |  |  |  | f$_0$(1710) 125±10 |
| mB$_2$X$_2$(1190) |  | π(1300) 200 to 600 |  | K*(1680) 322±110 |  | η(1760) 60±16 |

|  |  | a$_2$(1320) 104.7±1.9 |  |  |  |  |
|---|---|---|---|---|---|---|
| XXX(1260) |  | h$_1$(1380) 91±30 |  | K$_2$(1770) 186±14 |  | f$_2$(1810) 197±22 |
|  |  |  |  |  |  | **η(1820)** |
|  |  | π$_1$(1400) 300±40 |  | K$_3$*(1780) 159±21 |  | φ$_3$(1850) 87±28 |
| mX$_3$(1330) |  | a$_0$(1450) 265±13 |  | K$_2$(1820) 276±35 |  | η$_2$(1870) 225±14 |
|  |  | ρ(1450) 310±60 |  | K(1830) |  | f$_2$(1910) 163±50 |
| BX$_3$(1400) |  |  |  |  |  | f$_2$(1950) 452±22 |
| FX$_3$(1470) |  | π$_1$(1600) 312±64 |  | K$_0$*(1950) 201±113 |  | f$_2$(2010) 202±67 |
|  |  | a$_1$(1640) 300±62 |  | K$_2$*(1980) 373±93 |  | f$_0$(2020) 442±60 |
|  |  |  |  |  |  | f$_4$(2050) 194±13 |
| mFX$_3$(1540) |  | π$_2$(1670) 259±10 |  | K$_4$*(2045) 198±30 |  | f$_0$(2100) |
|  |  | ρ$_3$(1690) **161±10** |  |  |  |  |
|  |  | ρ(1700) **240±60** |  |  |  |  |
|  |  | a$_2$(1700) **256±40** |  |  |  |  |
| BFX$_3$(1610) |  |  |  |  |  | f$_2$(2150) **167±30** |
|  |  |  |  |  |  | f$_0$(2200) **201±51** |
| X$_4$(1680) |  | π(1800) **210±15** |  |  |  | f$_J$(2220) **23±8** |
|  |  |  |  |  |  | η(2225) |
| mX$_4$(1750) |  | ρ(1900) |  | K$_2$(2250) **180±30** |  | f$_2$(2300) **149±41** |
|  |  |  |  |  |  | f$_4$(2300) |
|  |  |  |  |  |  | f$_0$(2330) |
|  |  |  |  |  |  | f$_2$(2340) **319±81** |
| $BX_4(1820)$ | | $\rho_3(1990)$ | | $K_3(2320)$<br>**150±30** | | |
| $FX_4(1890)$ | | $a_4(2040)$<br>**360±40** | | $K_5^*(2380)$<br>**178±69** | | |
| $mFX_4(1960)$ | | $\pi_2(2100)$<br>**625±50** | | | | $f_6(2510)$<br>**255±40** |
| $BFX_4(2030)$ | | $\rho(2150)$<br>**363±50** | | $K_4(2500)$ | | |
| $X_5(2100)$ | | $\rho_3(2250)$ | | | | |
| $mX_5(2170)$ | | $\rho_5(2350)$<br>**400±100** | | | | |
| $FX_5(2310)$ | | $a_6(2450)$<br>**400±250** | | | | |